\documentclass[aps,prl,preprint,groupedaddress]{revtex4-1}

\usepackage{graphicx}

\begin{document}


\title{Multiferroic rust: magnetoelectric effect in g\"{o}thite, $\alpha$-FeOOH}


\author{N. V. Ter-Oganessian}
\email[]{nikita.teroganessian@gmail.com}
\affiliation{Institute of Physics, Southern Federal University, 344090 Rostov-on-Don, Russia}

\author{A. A. Guda}
\affiliation{International Research Center "Smart Materials", Southern Federal University, 344090 Rostov-on-Don, Russia}

\author{V. P. Sakhnenko}
\affiliation{Institute of Physics, Southern Federal University, 344090 Rostov-on-Don, Russia}


\date{\today}

\begin{abstract}
By means of symmetry analysis, density functional theory calculations, and Monte Carlo simulations we show that goethite, $\alpha$-FeOOH, is a linear magnetoelectric below its N\'{e}el temperature $T_{\rm N}=400$~K. The experimentally observed magnetic field induced spin-flop phase transition results in either change of direction of electric polarization or its suppression. Calculated value of magnetoelectric coefficient is 0.24~$\mu$C~m$^{-2}$~T$^{-1}$. The abundance of goethite in nature makes it arguably the most widespread magnetoelectric material.
\end{abstract}


\maketitle


The field of multiferroics has become one of the focal points in condensed matter physics during the last two decades. Mutual influence of magnetic and electric subsystems in magnetoelectrics opens up new opportunities for practical applications such as, for example, new types of logical elements, devices for storage of information, and various sensors~\cite{Scott_Review_2012,Ortega_Review_2015}. This stimulates search for new multiferroic materials both in the single-phase forms and as composites.  Recent advances in the physics and design of magnetoelectrics were summarized in numerous reviews (see, for example, Refs.~\cite{Dong_Review_2015,Young_Review_2015}).

Magnetoelectrics are known since late 1950's and were intensively studied during the last two decades. By now, many magnetoelectric (ME) crystals or even whole classes of such compounds are identified. However, the quest for new compounds continues due to the need for higher ME coupling constants and higher working temperatures.

Iron forms many oxides and hydroxides showing a plethora of magnetic properties, which also often develop at high temperatures~\cite{Cornel_Iron_Oxides_BOOK}. However, in contrast to, for example, chromium (Cr$_2$O$_3$)~\cite{Astrov_Cr2O3_1960}, cupric (CuO)~\cite{Kimura_CuO_2008}, or cobalt (Co$_3$O$_4$)~\cite{Saha_AB2O4_2016} oxides, only Fe$_3$O$_4$ was shown to display magnetoelectric properties~\cite{Kita_Fe3O4_1979,Yamauchi_Fe3O4_2009}.

Goethite, $\alpha$-FeOOH, is one of the most thermodynamically stable compounds out of iron oxides, hydroxides, or oxides-hydroxides, which arguably makes it the most abundant in nature among them~\cite{Cornel_Iron_Oxides_BOOK}. It is found in rocks and soils and is often responsible for their colour. In many parts of the world current climate favours mineralogical transformation of hematite ($\alpha$-Fe$_2$O$_3$) to goethite in soils and, therefore, the hematite-goethite ratio reflects the climate~\cite{Schwertmann_1971}. Goethite is also a common component of rusts, both atmospheric and electrochemical~\cite{Cornel_Iron_Oxides_BOOK}, and is found on Mars among other iron-containing minerals~\cite{Morris_Mars_2008}. In practical use goethite is an important pigment as it is a component of ochre deposits, however it also attracts interest in the form of suspensions of nanoparticles or nanorods showing considerable magnetic field-induced birefringence~\cite{Lemaire_birefrigence_2013,Li_birefrigence_2013}.

Here we show that goethite is linear magnetoelectric below its N\'{e}el temperature $T_{\rm N}=400$~K making it (i) a room temperature ME material, and (ii) arguably the most abundant ME material known to date. Using density functional theory (DFT) we identify the main exchange coupling constants of goethite and confirm its antiferromagnetic ground state, whereas Monte Carlo studies uncover its magnetoelectric behavior in magnetic fields.

\section{Results}

\begin{figure}
\includegraphics[height=7.8cm]{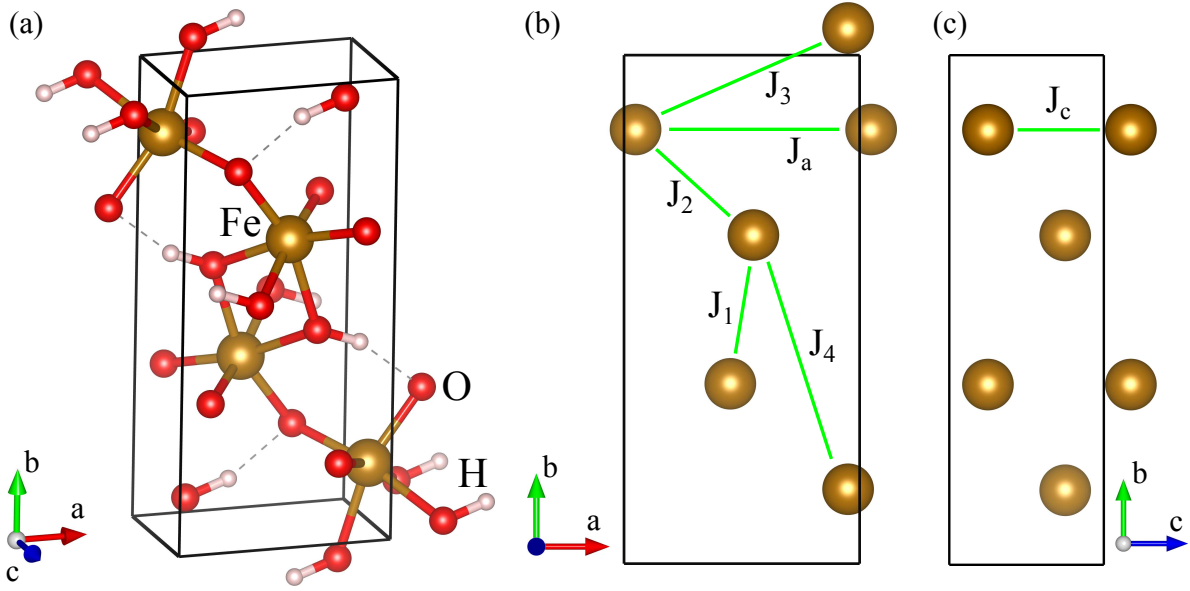}%
\caption{(a) Crystal structure of $\alpha$-FeOOH and (b-c) magnetic exchange paths \label{Fig:Fig1}}
\end{figure}
Goethite, $\alpha$-FeOOH, crystallizes in the orthorhombic structure with space group symmetry Pbnm (Z=4) shown in Fig.~\ref{Fig:Fig1}(a) and lattice parameters $a=4.5979$~\AA, $a=9.951$~\AA, and $c=3.0178$~\AA~\cite{Yang_2006}. Upon decreasing temperature it experiences an antiferromagnetic phase transition at temperature $T_{\rm N}$, which varies in the range from approximately 340 to 400~K depending on the purity of the sample~\cite{DeGrave_1986,Coey_1995,Bocquet_1995}. Below $T_{\rm N}$ the spins $\vec{S}_i$ of four iron ions Fe$_i$ ($i$=1, 2, 3, 4) located at positions $(0.0489,0.8537,1/4)$, $(0.9511,0.1463,3/4)$, $(0.5489,0.6463,3/4)$, and $(0.4511,0.3537,1/4)$~\cite{Yang_2006}, order antiferromagnetically with relative spin arrangement $(+--+)$, respectively~\cite{Forsyth_1968,Zepeda_Alarcon_2014}. This ordered spin arrangement can be described by the order parameter $\vec{A}$. Other possible spin arrangements with $\vec{k}=0$ described by the order parameters $\vec{F}$, $\vec{G}$, and $\vec{C}$ are summarized in Table~\ref{tab:IRs}. The direction of the ordered spins is experimentally found to be along the $c$ axis of the crystal cell. Therefore, the appearing magnetic structure with the wave vector $\vec{k}=0$ can be described by the order parameter $A_z$. Below we adopt an orthogonal system of axes $x$, $y$, and $z$ being parallel to the crystal axes $a$, $b$, and $c$, respectively.
\begin{table}
\caption{Spin arrangements of the Fe$_i$ ions with $\vec{k}=0$. First four columns give relative spin orderings of Fe$_i$ spins. The last column gives the irreducible representations (IR) according to which the components $x$, $y$, and $z$ of the order parameters transform, respectively.\label{tab:IRs}}
\begin{ruledtabular}
\begin{tabular}{llllcc}
Fe$_1$ & Fe$_2$ & Fe$_3$ & Fe$_4$ & Order parameter & IR's\\
\hline
$+$ & $+$ & $+$ & $+$ & $\vec{F}$ & $\Gamma^{2+}$, $\Gamma^{3+}$, $\Gamma^{4+}$ \\
$+$ & $-$ & $+$ & $-$ & $\vec{G}$ & $\Gamma^{1-}$, $\Gamma^{4-}$, $\Gamma^{3-}$ \\
$+$ & $+$ & $-$ & $-$ & $\vec{C}$ & $\Gamma^{3+}$, $\Gamma^{2+}$, $\Gamma^{1+}$ \\
$+$ & $-$ & $-$ & $+$ & $\vec{A}$ & $\Gamma^{4-}$, $\Gamma^{1-}$, $\Gamma^{2-}$ \\
\end{tabular}
\end{ruledtabular}
\end{table}

The symmetry of magnetic structure with $A_z\neq0$ appearing below T$_N$ is Pb$'$nm~\cite{Zepeda_Alarcon_2014} and allows linear magnetoelectric effect with magnetoelectric interactions given by
\begin{eqnarray}
A_zF_yP_z,\label{eq:invariant_AzFyPz}\\
A_zF_zP_y,\label{eq:invariant_AzFzPy}
\end{eqnarray}
where $\vec{F}$ and $\vec{P}$ are ferromagnetic moment and electric polarization, respectively. Thus, in the antiferromagnetic phase magnetic field applied along the $y$ or $z$ axis induces electric polarization components $P_z$ or $P_y$, respectively. It is found, however, that sufficiently strong magnetic field along the $z$ axis results in a spin-flop transition, in which the spins reorient towards either the $x$ or the $y$ axis~\cite{Coey_1995}. This will be discussed in more detail below.

It has to be noted here, that in the case when the initial paraelectric and paramagnetic phase possesses inversion symmetry operation, a magnetic phase transition with $\vec{k}=0$ occurring according to a single irreducible representation cannot induce electric polarization~\cite{Kovalev_1973}. However, linear magnetoelectric effect can be possible, as is the case in $\alpha$-FeOOH: when $A_z\neq0$ appears, the inversion symmetry is broken, but spatial inversion together with time reversal operation is a symmetry element, which results in interactions~(\ref{eq:invariant_AzFyPz}) and~(\ref{eq:invariant_AzFzPy}).

Using density functional theory we calculate six magnetic exchange constants, which are summarized in Table~\ref{tab:Js} and the respective exchange paths are shown in Figs.~\ref{Fig:Fig1}(b-c).
\begin{table}
\caption{Calculated magnetic exchange constants for $\alpha$-FeOOH in meV.\label{tab:Js}}
\begin{ruledtabular}
\begin{tabular}{ccccccc}
 & J$_1$ & J$_2$ & J$_3$ & J$_4$ & J$_a$ & J$_c$ \\
\hline
Fe -- Fe distance, \AA & 3.310 & 3.438 & 5.288 & 5.308 & 4.598 & 3.018 \\
J, meV & 15.1 & 48.1 & -0.28 & 3.19 & 4.4 & 17.7 \\
\end{tabular}
\end{ruledtabular}
\end{table}
It is found that the exchange couplings are mostly antiferromagnetic and the magnetic ground state is described by $\vec{A}\neq0$ in accordance with the experiments.

\begin{figure}
\includegraphics[height=9.8cm]{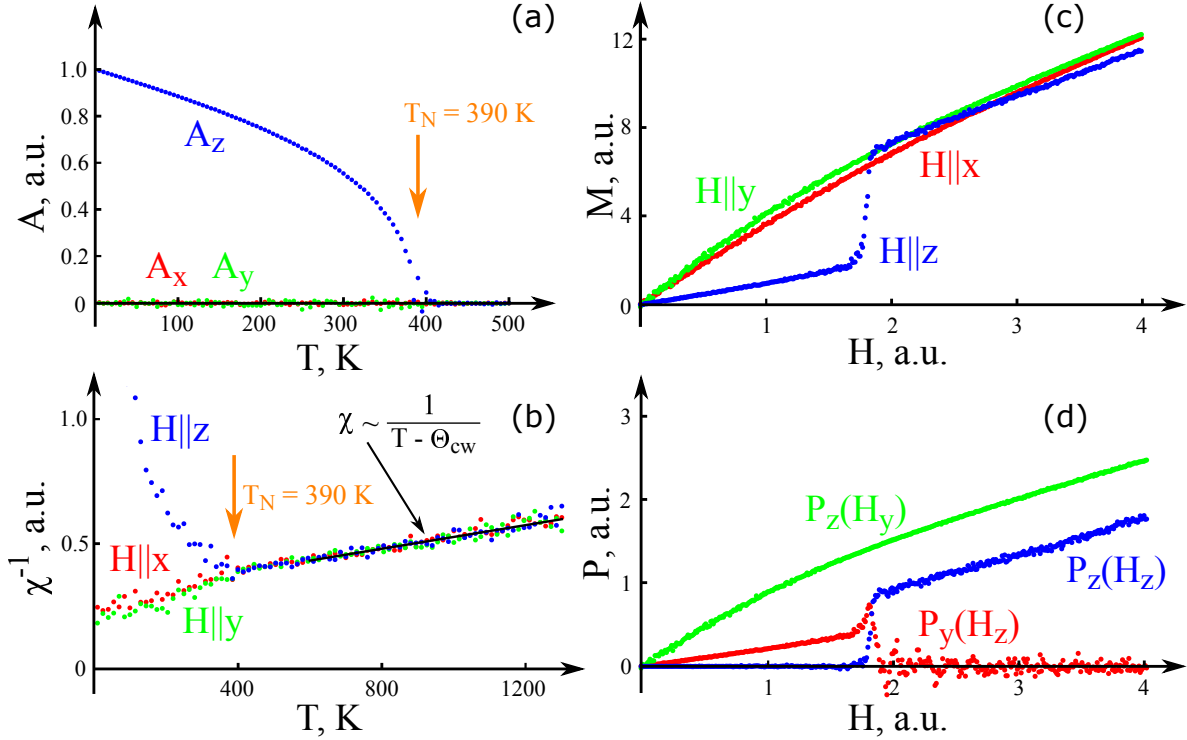}%
\caption{Results of Monte Carlo calculations. (a) temperature dependence of the order parameters $A_x$, $A_y$, and $A_z$. (b) reciprocal magnetic susceptibility for various directions as function of temperature and a fit with the Curie-Weiss law (solid line). (c) magnetization and (d) electric polarization at $T=100$~K as function of magnetic field.\label{Fig:Fig2}}
\end{figure}

Monte Carlo calculations reveal that with the found exchange constants the N\'{e}el temperature $T_{\rm N}^{\rm MC}=390$~K is slightly lower than in experiments. Figure~\ref{Fig:Fig2}(a) shows temperature dependence of the order parameters, revealing that $A_z$ emerges at $T_{\rm N}^{\rm MC}$ confirming the appearance of antiferromagnetic order. The fit of magnetic susceptibility in the paramagnetic region by $\chi=C/(T-\Theta_{\rm CW})$ shown in Fig.~\ref{Fig:Fig2}(b) gives the Curie-Weiss temperature $\Theta_{\rm CW}\approx-1250$~K. This implies that in goethite considerable magnetic frustration exists since $|\Theta_{\rm CW}|/T_{\rm N}\approx3.2$.

At $H_c=20$~T a spin-flop transition occurs in goethite~\cite{Coey_1995} resulting in rotation of the antiferromagnetic vector to either $a$- or $b$-axis. Our results on magnetic field dependence of magnetization in the antiferromagnetic phase shown in Fig.~\ref{Fig:Fig2}(c) are in qualitative agreement with the experimental data~\cite{Coey_1995}. In our Monte Carlo simulations we assume $D_x>D_y$, which results in appearance of $A_y$ at $H_z\gtrsim1.8$~a.u. and corresponding vanishing of $A_z$.

Figure~\ref{Fig:Fig2}(d) shows $H$-dependence of electric polarization calculated using Eqs.~(\ref{eq:invariant_AzFyPz}) and~(\ref{eq:invariant_AzFzPy}) and the ME interaction
\begin{equation}
A_yF_zP_z,\label{eq:invariant_AyFzPz}
\end{equation}
which is relevant in the spin-flopped phase in the case when $D_x>D_y$. It follows that in the antiferromagnetic phase $\alpha$-FeOOH is a linear magnetoelectric, since external $H_y$ and $H_z$ induce $P_z$ and $P_y$, respectively. Furthermore, at $H_z=20$~T a flop of polarization from the $b$- to $c$-axis may occur. In the case $D_x<D_y$ the antiferromagnetic vector will flip to $A_x$ at $H_c\gtrsim20$~T resulting in disappearance of electric polarization.

The microscopic origin of ME effect can be understood rewriting the ME interaction~(\ref{eq:invariant_AzFyPz}) through spins
\[
I_1 = A_zF_yP_z = w_1+w_2+w_3-w_4,
\]
where
\begin{eqnarray}
w_1=P_z ( S_{1y} S_{1z} - S_{2y} S_{2z} - S_{3y} S_{3z} + S_{4y} S_{4z}),\nonumber\\
w_2=P_z ( S_{1z} S_{2y} - S_{1y} S_{2z} - S_{3z} S_{4y} + S_{3y} S_{4z}),\nonumber\\
w_3=P_z ( S_{1z} S_{3y} - S_{1y} S_{3z} - S_{2z} S_{4y} + S_{2y} S_{4z}),\nonumber\\
w_4=P_z ( S_{2z} S_{3y} + S_{2y} S_{3z} - S_{1z} S_{4y} - S_{1y} S_{4z}).\nonumber
\end{eqnarray}
The interaction $w_1$ is a single-ion contribution, whereas $w_2$, $w_3$, and $w_4$ result from interactions of two spins. Thus, the ME coupling may have both single-ion and two-ion contributions. The single-ion contribution is in accordance with the local non-centrosymmetric crystal environment of Fe atoms, the local crystal symmetry of which is a mirror plane $\sigma_z$ oriented parallel to the $xy$ plane. Thus, it allows local spin-dependent electric dipole moments of electron orbitals $d_z\sim S_yS_z$~\cite{Sakhnenko_2012}.

In order to estimate the value of magnetically induced electric polarization we performed non-collinear DFT calculations. The spins were first relaxed in the stable $A_z$ configuration and then constrained to give additional ferromagnetic component $F_y$. Artificially induced ferromagnetic ordering amounted to approximately 0.62~$\mu_{\rm B}$ per f.u. and resulted in rotation of spins away from the $z$ axis by about 8.8$^\circ$. The resulting electric polarization calculated using the Berry phase approach was found to be 49.3~$\mu$C/m$^2$. Taking the experimental magnetic susceptibility of approximately 0.003~$\mu_{\rm B}$/T per Fe$^{3+}$ ion~\cite{Martin_Hernandez_2010,Pankhurst_2012}  we can estimate the ME coefficient to be of the order of 0.24~$\mu$C~m$^{-2}$~T$^{-1}$, which is comparable to that of LiNiPO$_4$~\cite{Sakhnenko_2012,Kornev_2000}.

Relative values of different contributions to ME effect can be estimated from DFT calculations. For this purpose one can use the ME interactions
\begin{eqnarray}
I_2 = C_zG_yP_z = w_1-w_2+w_3+w_4,\nonumber\\
I_3 = A_yF_zP_z = w_1-w_2-w_3-w_4,\nonumber\\
I_4 = C_yG_zP_z = w_1+w_2-w_3+w_4.\nonumber
\end{eqnarray}
Performing calculations using the magnetic configurations $C_zG_y$, $A_yF_z$, and $C_yG_z$ similar to above and evaluating $P_z$ using the Berry phase approach we find that the biggest contribution to ME effect is $w_3$ and the other contributions relative to $w_3$ are $w_1/w_3\approx-0.034$, $w_2/w_3\approx-0.21$, and $w_4/w_3=0$. Therefore, it follows that $w_1$ and $w_2$ act in the direction opposite to $w_3$.

\section{Conclusions}

Based on the symmetry analysis of the available crystal and magnetic structures of goethite, $\alpha$-FeOOH, we suggest that it is linear magnetoelectric below its N\'{e}el temperature. Using density functional calculations and Monte Carlo simulations we find main exchange constants in goethite and calculate its magnetic and magnetoelectric behavior.

Goethite belongs to the $\alpha$-AlOOH diaspore structural type, which is also shared by, for example, $\alpha$-MnOOH, Fe(OH)F, and Co(OH)F. The latter compound is also antiferromagnetic below $\sim40$~K with the spin arrangement similar to $\alpha$-FeOOH~\cite{Yahia_CoOHF_2014} and should, thus, display similar linear ME properties below its $T_{\rm N}$.

Nature creates beautiful polycrystalline goethite samples, which are encountered in significant amounts in various deposits. However, synthesis of single crystals in laboratory or preparation of good ceramic samples can be a challenge, as $\alpha$-FeOOH starts to decompose at temperatures higher than 200~$^\circ$C to form hematite, $\alpha$-Fe$_2$O$_3$. In this respect it may be easier to show the magnetoelectric behavior experimentally in the aforementioned isostructural compounds with similar magnetic structure, e.g., in Co(OH)F.

\section{Methods}

\textbf{DFT calculations.} Density functional theory calculations were performed using the Vienna {\it Ab-initio} Simulation Package (VASP)~\cite{Kresse_1996} and the projected augmented wave method~\cite{Bloechl_1994}. We used the GGA exchange correlation approximation corrected by means of the GGA+U formalism for the Fe atoms with $U_{\rm eff}=U-J=3$~eV within the Dudarev approach~\cite{Dudarev_1998}. This value of $U_{\rm eff}$ was shown earlier to properly account for the structural and magnetic properties of $\alpha$-FeOOH~\cite{Tunega_DFT_2012,Meng_When_meets_oxides_2016}. The energy cutoff was 500~eV, whereas the Brillouin zone integration was done using the 8 x 4 x 12 set of $k$-points determined by the Monkhorst-Pack scheme. The calculated lattice parameters $a=4.638$~\AA, $b=10.037$~\AA, and $c=3.038$~\AA~are within 1\%
of the experimentally determined values~\cite{Zepeda_Alarcon_2014,Yang_2006}. The local magnetic moment value of 4.14~$\mu_{\rm B}$ of Fe ions is between the experimentally reported values of 3.8~$\mu_{\rm B}$~\cite{Bocquet_1992} and 4.45~$\mu_{\rm B}$~\cite{Zepeda_Alarcon_2014}. The band gap 1.9~eV obtained in DFT calculations is slightly lower than the experimental values 2.1 -- 2.5~eV~\cite{Leland_1987,Sherman_2005}.
Electric polarization was calculated using the Berry phase approach as implemented in VASP.

\textbf{Classical Monte-Carlo simulations.} Classical Monte Carlo simulations using the exchange constants determined by DFT calculations are performed using the Hamiltonian
\[
\mathcal{H}=\sum_{ij}J_{ij}\vec{S}_i\cdot\vec{S}_j+\sum_{i}\left(
D_xS_{ix}^2+
D_yS_{iy}^2+
D_zS_{iz}^2\right)-\vec{H}\cdot\vec{S},
\]
where $\vec{S}$ are classical vectors of unit length, $D_\alpha$ ($\alpha=x,y,z$) are anisotropy constants, and $\vec{H}$ is magnetic field. The calculations are performed using the Metropolis scheme and a simulation box with dimensions $12\times12\times12$ unit cells. In our simulations we tentatively use $D_x=-D_z=1.5$~eV and $D_y=0$, which reflects the easy axis direction parallel to the $c$-axis.

\section{Acknowledgements}

N.V.T. acknowledges financial support by the RA MES SCS, within the frames of the ``RA MES SCS - YSU - RF SFEDU'' international call for joint project No. VnGr-07/2017-32.

\section{Author Contributions}

N.V.T. conceived the project. A.A.G. performed DFT calculations. N.V.T. did the Monte Carlo simulations. N.V.T. and V.P.S. supervised the research and wrote the manuscript. All authors discussed the results and commented the manuscript.

\section{Additional information}

Competing financial interests: The authors declare no competing financial interests.


\begin{thebibliography}{34}%
\makeatletter
\providecommand \@ifxundefined [1]{%
 \@ifx{#1\undefined}
}%
\providecommand \@ifnum [1]{%
 \ifnum #1\expandafter \@firstoftwo
 \else \expandafter \@secondoftwo
 \fi
}%
\providecommand \@ifx [1]{%
 \ifx #1\expandafter \@firstoftwo
 \else \expandafter \@secondoftwo
 \fi
}%
\providecommand \natexlab [1]{#1}%
\providecommand \enquote  [1]{``#1''}%
\providecommand \bibnamefont  [1]{#1}%
\providecommand \bibfnamefont [1]{#1}%
\providecommand \citenamefont [1]{#1}%
\providecommand \href@noop [0]{\@secondoftwo}%
\providecommand \href [0]{\begingroup \@sanitize@url \@href}%
\providecommand \@href[1]{\@@startlink{#1}\@@href}%
\providecommand \@@href[1]{\endgroup#1\@@endlink}%
\providecommand \@sanitize@url [0]{\catcode `\\12\catcode `\$12\catcode
  `\&12\catcode `\#12\catcode `\^12\catcode `\_12\catcode `\%12\relax}%
\providecommand \@@startlink[1]{}%
\providecommand \@@endlink[0]{}%
\providecommand \url  [0]{\begingroup\@sanitize@url \@url }%
\providecommand \@url [1]{\endgroup\@href {#1}{\urlprefix }}%
\providecommand \urlprefix  [0]{URL }%
\providecommand \Eprint [0]{\href }%
\providecommand \doibase [0]{http://dx.doi.org/}%
\providecommand \selectlanguage [0]{\@gobble}%
\providecommand \bibinfo  [0]{\@secondoftwo}%
\providecommand \bibfield  [0]{\@secondoftwo}%
\providecommand \translation [1]{[#1]}%
\providecommand \BibitemOpen [0]{}%
\providecommand \bibitemStop [0]{}%
\providecommand \bibitemNoStop [0]{.\EOS\space}%
\providecommand \EOS [0]{\spacefactor3000\relax}%
\providecommand \BibitemShut  [1]{\csname bibitem#1\endcsname}%
\let\auto@bib@innerbib\@empty
\bibitem [{\citenamefont {Scott}(2012)}]{Scott_Review_2012}%
  \BibitemOpen
  \bibfield  {author} {\bibinfo {author} {\bibfnamefont {J.~F.}\ \bibnamefont
  {Scott}},\ }\href {\doibase 10.1039/C2JM16137K} {\bibfield  {journal}
  {\bibinfo  {journal} {J. Mater. Chem.}\ }\textbf {\bibinfo {volume} {22}},\
  \bibinfo {pages} {4567} (\bibinfo {year} {2012})}\BibitemShut {NoStop}%
\bibitem [{\citenamefont {Ortega}\ \emph {et~al.}(2015)\citenamefont {Ortega},
  \citenamefont {Kumar}, \citenamefont {Scott},\ and\ \citenamefont
  {Katiyar}}]{Ortega_Review_2015}%
  \BibitemOpen
  \bibfield  {author} {\bibinfo {author} {\bibfnamefont {N.}~\bibnamefont
  {Ortega}}, \bibinfo {author} {\bibfnamefont {A.}~\bibnamefont {Kumar}},
  \bibinfo {author} {\bibfnamefont {J.~F.}\ \bibnamefont {Scott}}, \ and\
  \bibinfo {author} {\bibfnamefont {R.~S.}\ \bibnamefont {Katiyar}},\ }\href
  {\doibase 10.1088/0953-8984/27/50/504002} {\bibfield  {journal} {\bibinfo
  {journal} {J. Phys.: Condens. Matter}\ }\textbf {\bibinfo {volume} {27}},\
  \bibinfo {pages} {504002} (\bibinfo {year} {2015})}\BibitemShut {NoStop}%
\bibitem [{\citenamefont {Dong}\ \emph {et~al.}(2015)\citenamefont {Dong},
  \citenamefont {Liu}, \citenamefont {Cheong},\ and\ \citenamefont
  {Ren}}]{Dong_Review_2015}%
  \BibitemOpen
  \bibfield  {author} {\bibinfo {author} {\bibfnamefont {S.}~\bibnamefont
  {Dong}}, \bibinfo {author} {\bibfnamefont {J.-M.}\ \bibnamefont {Liu}},
  \bibinfo {author} {\bibfnamefont {S.-W.}\ \bibnamefont {Cheong}}, \ and\
  \bibinfo {author} {\bibfnamefont {Z.}~\bibnamefont {Ren}},\ }\href {\doibase
  10.1080/00018732.2015.1114338} {\bibfield  {journal} {\bibinfo  {journal}
  {Adv. Phys.}\ }\textbf {\bibinfo {volume} {64}},\ \bibinfo {pages} {519}
  (\bibinfo {year} {2015})}\BibitemShut {NoStop}%
\bibitem [{\citenamefont {Young}\ \emph {et~al.}(2015)\citenamefont {Young},
  \citenamefont {Stroppa}, \citenamefont {Picozzi},\ and\ \citenamefont
  {Rondinelli}}]{Young_Review_2015}%
  \BibitemOpen
  \bibfield  {author} {\bibinfo {author} {\bibfnamefont {J.}~\bibnamefont
  {Young}}, \bibinfo {author} {\bibfnamefont {A.}~\bibnamefont {Stroppa}},
  \bibinfo {author} {\bibfnamefont {S.}~\bibnamefont {Picozzi}}, \ and\
  \bibinfo {author} {\bibfnamefont {J.~M.}\ \bibnamefont {Rondinelli}},\ }\href
  {\doibase 10.1088/0953-8984/27/28/283202} {\bibfield  {journal} {\bibinfo
  {journal} {J. Phys.: Condens. Matter}\ }\textbf {\bibinfo {volume} {27}},\
  \bibinfo {pages} {283202} (\bibinfo {year} {2015})}\BibitemShut {NoStop}%
\bibitem [{\citenamefont {Cornell}\ and\ \citenamefont
  {Schwertmann}(2003)}]{Cornel_Iron_Oxides_BOOK}%
  \BibitemOpen
  \bibfield  {author} {\bibinfo {author} {\bibfnamefont {R.~M.}\ \bibnamefont
  {Cornell}}\ and\ \bibinfo {author} {\bibfnamefont {U.}~\bibnamefont
  {Schwertmann}},\ }\href@noop {} {\emph {\bibinfo {title} {The Iron Oxides:
  Structure, Properties, Reactions, Occurences and Uses}}}\ (\bibinfo
  {publisher} {WILEY-VCH Verlag, Weinheim},\ \bibinfo {year}
  {2003})\BibitemShut {NoStop}%
\bibitem [{\citenamefont {Astrov}(1960)}]{Astrov_Cr2O3_1960}%
  \BibitemOpen
  \bibfield  {author} {\bibinfo {author} {\bibfnamefont {D.~N.}\ \bibnamefont
  {Astrov}},\ }\href@noop {} {\bibfield  {journal} {\bibinfo  {journal} {JETP}\
  }\textbf {\bibinfo {volume} {11}},\ \bibinfo {pages} {708} (\bibinfo {year}
  {1960})}\BibitemShut {NoStop}%
\bibitem [{\citenamefont {Kimura}\ \emph {et~al.}(2008)\citenamefont {Kimura},
  \citenamefont {Sekio}, \citenamefont {Nakamura}, \citenamefont {Siegrist},\
  and\ \citenamefont {Ramirez}}]{Kimura_CuO_2008}%
  \BibitemOpen
  \bibfield  {author} {\bibinfo {author} {\bibfnamefont {T.}~\bibnamefont
  {Kimura}}, \bibinfo {author} {\bibfnamefont {Y.}~\bibnamefont {Sekio}},
  \bibinfo {author} {\bibfnamefont {H.}~\bibnamefont {Nakamura}}, \bibinfo
  {author} {\bibfnamefont {T.}~\bibnamefont {Siegrist}}, \ and\ \bibinfo
  {author} {\bibfnamefont {A.~P.}\ \bibnamefont {Ramirez}},\ }\href {\doibase
  10.1038/nmat2125} {\bibfield  {journal} {\bibinfo  {journal} {Nat. Mater.}\
  }\textbf {\bibinfo {volume} {7}},\ \bibinfo {pages} {291} (\bibinfo {year}
  {2008})}\BibitemShut {NoStop}%
\bibitem [{\citenamefont {Saha}\ \emph {et~al.}(2016)\citenamefont {Saha},
  \citenamefont {Ghara}, \citenamefont {Suard}, \citenamefont {Jang},
  \citenamefont {Kim}, \citenamefont {Ter-Oganessian},\ and\ \citenamefont
  {Sundaresan}}]{Saha_AB2O4_2016}%
  \BibitemOpen
  \bibfield  {author} {\bibinfo {author} {\bibfnamefont {R.}~\bibnamefont
  {Saha}}, \bibinfo {author} {\bibfnamefont {S.}~\bibnamefont {Ghara}},
  \bibinfo {author} {\bibfnamefont {E.}~\bibnamefont {Suard}}, \bibinfo
  {author} {\bibfnamefont {D.~H.}\ \bibnamefont {Jang}}, \bibinfo {author}
  {\bibfnamefont {K.~H.}\ \bibnamefont {Kim}}, \bibinfo {author} {\bibfnamefont
  {N.~V.}\ \bibnamefont {Ter-Oganessian}}, \ and\ \bibinfo {author}
  {\bibfnamefont {A.}~\bibnamefont {Sundaresan}},\ }\href {\doibase
  10.1103/PhysRevB.94.014428} {\bibfield  {journal} {\bibinfo  {journal} {Phys.
  Rev. B}\ }\textbf {\bibinfo {volume} {94}},\ \bibinfo {pages} {014428}
  (\bibinfo {year} {2016})}\BibitemShut {NoStop}%
\bibitem [{\citenamefont {Kita}\ \emph {et~al.}(1979)\citenamefont {Kita},
  \citenamefont {Siratori}, \citenamefont {Kohn}, \citenamefont {Tasaki},
  \citenamefont {Kimura},\ and\ \citenamefont {Shindo}}]{Kita_Fe3O4_1979}%
  \BibitemOpen
  \bibfield  {author} {\bibinfo {author} {\bibfnamefont {E.}~\bibnamefont
  {Kita}}, \bibinfo {author} {\bibfnamefont {K.}~\bibnamefont {Siratori}},
  \bibinfo {author} {\bibfnamefont {K.}~\bibnamefont {Kohn}}, \bibinfo {author}
  {\bibfnamefont {A.}~\bibnamefont {Tasaki}}, \bibinfo {author} {\bibfnamefont
  {S.}~\bibnamefont {Kimura}}, \ and\ \bibinfo {author} {\bibfnamefont
  {I.}~\bibnamefont {Shindo}},\ }\href {\doibase 10.1143/JPSJ.47.1788}
  {\bibfield  {journal} {\bibinfo  {journal} {J. Phys. Soc. Jpn.}\ }\textbf
  {\bibinfo {volume} {47}},\ \bibinfo {pages} {1788} (\bibinfo {year}
  {1979})}\BibitemShut {NoStop}%
\bibitem [{\citenamefont {Yamauchi}\ \emph {et~al.}(2009)\citenamefont
  {Yamauchi}, \citenamefont {Fukushima},\ and\ \citenamefont
  {Picozzi}}]{Yamauchi_Fe3O4_2009}%
  \BibitemOpen
  \bibfield  {author} {\bibinfo {author} {\bibfnamefont {K.}~\bibnamefont
  {Yamauchi}}, \bibinfo {author} {\bibfnamefont {T.}~\bibnamefont {Fukushima}},
  \ and\ \bibinfo {author} {\bibfnamefont {S.}~\bibnamefont {Picozzi}},\ }\href
  {\doibase 10.1103/PhysRevB.79.212404} {\bibfield  {journal} {\bibinfo
  {journal} {Phys. Rev. B}\ }\textbf {\bibinfo {volume} {79}},\ \bibinfo
  {pages} {212404} (\bibinfo {year} {2009})}\BibitemShut {NoStop}%
\bibitem [{\citenamefont {Schwertmann}(1971)}]{Schwertmann_1971}%
  \BibitemOpen
  \bibfield  {author} {\bibinfo {author} {\bibfnamefont {U.}~\bibnamefont
  {Schwertmann}},\ }\href {\doibase 10.1038/232624a0} {\bibfield  {journal}
  {\bibinfo  {journal} {Nature}\ }\textbf {\bibinfo {volume} {232}},\ \bibinfo
  {pages} {624} (\bibinfo {year} {1971})}\BibitemShut {NoStop}%
\bibitem [{\citenamefont {Morris}\ \emph {et~al.}(2008)\citenamefont {Morris},
  \citenamefont {Klingelh\"{o}fer}, \citenamefont {Schr\"{o}der}, \citenamefont
  {Fleischer}, \citenamefont {Ming}, \citenamefont {Yen}, \citenamefont
  {Gellert}, \citenamefont {Arvidson}, \citenamefont {Rodionov}, \citenamefont
  {Crumpler}, \citenamefont {Clark}, \citenamefont {Cohen}, \citenamefont
  {McCoy}, \citenamefont {Mittlefehldt}, \citenamefont {Schmidt}, \citenamefont
  {{de Souza}},\ and\ \citenamefont {Squyres}}]{Morris_Mars_2008}%
  \BibitemOpen
  \bibfield  {author} {\bibinfo {author} {\bibfnamefont {R.~V.}\ \bibnamefont
  {Morris}}, \bibinfo {author} {\bibfnamefont {G.}~\bibnamefont
  {Klingelh\"{o}fer}}, \bibinfo {author} {\bibfnamefont {C.}~\bibnamefont
  {Schr\"{o}der}}, \bibinfo {author} {\bibfnamefont {I.}~\bibnamefont
  {Fleischer}}, \bibinfo {author} {\bibfnamefont {D.~W.}\ \bibnamefont {Ming}},
  \bibinfo {author} {\bibfnamefont {A.~S.}\ \bibnamefont {Yen}}, \bibinfo
  {author} {\bibfnamefont {R.}~\bibnamefont {Gellert}}, \bibinfo {author}
  {\bibfnamefont {R.~E.}\ \bibnamefont {Arvidson}}, \bibinfo {author}
  {\bibfnamefont {D.~S.}\ \bibnamefont {Rodionov}}, \bibinfo {author}
  {\bibfnamefont {L.~S.}\ \bibnamefont {Crumpler}}, \bibinfo {author}
  {\bibfnamefont {B.~C.}\ \bibnamefont {Clark}}, \bibinfo {author}
  {\bibfnamefont {B.~A.}\ \bibnamefont {Cohen}}, \bibinfo {author}
  {\bibfnamefont {T.~J.}\ \bibnamefont {McCoy}}, \bibinfo {author}
  {\bibfnamefont {D.~W.}\ \bibnamefont {Mittlefehldt}}, \bibinfo {author}
  {\bibfnamefont {M.~E.}\ \bibnamefont {Schmidt}}, \bibinfo {author}
  {\bibfnamefont {P.~A.}\ \bibnamefont {{de Souza}}}, \ and\ \bibinfo {author}
  {\bibfnamefont {S.~W.}\ \bibnamefont {Squyres}},\ }\href {\doibase
  10.1029/2008JE003201} {\bibfield  {journal} {\bibinfo  {journal} {J. Geophys.
  Res.}\ }\textbf {\bibinfo {volume} {113}},\ \bibinfo {pages} {E12S42}
  (\bibinfo {year} {2008})}\BibitemShut {NoStop}%
\bibitem [{\citenamefont {Lemaire}\ \emph {et~al.}(2002)\citenamefont
  {Lemaire}, \citenamefont {Davidson}, \citenamefont {Ferr\'{e}}, \citenamefont
  {Jamet}, \citenamefont {Panine}, \citenamefont {Dozov},\ and\ \citenamefont
  {Jolivet}}]{Lemaire_birefrigence_2013}%
  \BibitemOpen
  \bibfield  {author} {\bibinfo {author} {\bibfnamefont {B.~J.}\ \bibnamefont
  {Lemaire}}, \bibinfo {author} {\bibfnamefont {P.}~\bibnamefont {Davidson}},
  \bibinfo {author} {\bibfnamefont {J.}~\bibnamefont {Ferr\'{e}}}, \bibinfo
  {author} {\bibfnamefont {J.~P.}\ \bibnamefont {Jamet}}, \bibinfo {author}
  {\bibfnamefont {P.}~\bibnamefont {Panine}}, \bibinfo {author} {\bibfnamefont
  {I.}~\bibnamefont {Dozov}}, \ and\ \bibinfo {author} {\bibfnamefont {J.~P.}\
  \bibnamefont {Jolivet}},\ }\href {\doibase 10.1103/PhysRevLett.88.125507}
  {\bibfield  {journal} {\bibinfo  {journal} {Phys. Rev. Lett.}\ }\textbf
  {\bibinfo {volume} {88}},\ \bibinfo {pages} {125507} (\bibinfo {year}
  {2002})}\BibitemShut {NoStop}%
\bibitem [{\citenamefont {Li}\ \emph {et~al.}(2013)\citenamefont {Li},
  \citenamefont {Qiu}, \citenamefont {Lin}, \citenamefont {Chen}, \citenamefont
  {Liu},\ and\ \citenamefont {Li}}]{Li_birefrigence_2013}%
  \BibitemOpen
  \bibfield  {author} {\bibinfo {author} {\bibfnamefont {J.}~\bibnamefont
  {Li}}, \bibinfo {author} {\bibfnamefont {X.}~\bibnamefont {Qiu}}, \bibinfo
  {author} {\bibfnamefont {Y.}~\bibnamefont {Lin}}, \bibinfo {author}
  {\bibfnamefont {L.}~\bibnamefont {Chen}}, \bibinfo {author} {\bibfnamefont
  {X.}~\bibnamefont {Liu}}, \ and\ \bibinfo {author} {\bibfnamefont
  {D.}~\bibnamefont {Li}},\ }\href {\doibase 10.1016/j.cplett.2013.10.082}
  {\bibfield  {journal} {\bibinfo  {journal} {Chem. Phys. Lett.}\ }\textbf
  {\bibinfo {volume} {590}},\ \bibinfo {pages} {168} (\bibinfo {year}
  {2013})}\BibitemShut {NoStop}%
\bibitem [{\citenamefont {Yang}\ \emph {et~al.}(2006)\citenamefont {Yang},
  \citenamefont {Lu}, \citenamefont {Downs},\ and\ \citenamefont
  {Costin}}]{Yang_2006}%
  \BibitemOpen
  \bibfield  {author} {\bibinfo {author} {\bibfnamefont {H.}~\bibnamefont
  {Yang}}, \bibinfo {author} {\bibfnamefont {R.}~\bibnamefont {Lu}}, \bibinfo
  {author} {\bibfnamefont {R.~T.}\ \bibnamefont {Downs}}, \ and\ \bibinfo
  {author} {\bibfnamefont {G.}~\bibnamefont {Costin}},\ }\href {\doibase
  10.1107/S1600536806047258} {\bibfield  {journal} {\bibinfo  {journal} {Acta
  Cryst. E}\ }\textbf {\bibinfo {volume} {62}},\ \bibinfo {pages} {i250}
  (\bibinfo {year} {2006})}\BibitemShut {NoStop}%
\bibitem [{\citenamefont {{De Grave}}\ and\ \citenamefont
  {Vandenberghe}(1986)}]{DeGrave_1986}%
  \BibitemOpen
  \bibfield  {author} {\bibinfo {author} {\bibfnamefont {E.}~\bibnamefont {{De
  Grave}}}\ and\ \bibinfo {author} {\bibfnamefont {R.~E.}\ \bibnamefont
  {Vandenberghe}},\ }\href {\doibase 10.1007/BF02061530} {\bibfield  {journal}
  {\bibinfo  {journal} {Hyper. Interactions}\ }\textbf {\bibinfo {volume}
  {28}},\ \bibinfo {pages} {643} (\bibinfo {year} {1986})}\BibitemShut
  {NoStop}%
\bibitem [{\citenamefont {Coey}\ \emph {et~al.}(1995)\citenamefont {Coey},
  \citenamefont {Barry}, \citenamefont {Brotto}, \citenamefont {Rakoto},
  \citenamefont {Brennan}, \citenamefont {Mussel}, \citenamefont {Collomb},\
  and\ \citenamefont {Fruchart}}]{Coey_1995}%
  \BibitemOpen
  \bibfield  {author} {\bibinfo {author} {\bibfnamefont {J.~M.~D.}\
  \bibnamefont {Coey}}, \bibinfo {author} {\bibfnamefont {A.}~\bibnamefont
  {Barry}}, \bibinfo {author} {\bibfnamefont {J.}~\bibnamefont {Brotto}},
  \bibinfo {author} {\bibfnamefont {H.}~\bibnamefont {Rakoto}}, \bibinfo
  {author} {\bibfnamefont {S.}~\bibnamefont {Brennan}}, \bibinfo {author}
  {\bibfnamefont {W.~N.}\ \bibnamefont {Mussel}}, \bibinfo {author}
  {\bibfnamefont {A.}~\bibnamefont {Collomb}}, \ and\ \bibinfo {author}
  {\bibfnamefont {D.}~\bibnamefont {Fruchart}},\ }\href {\doibase
  10.1088/0953-8984/7/4/006} {\bibfield  {journal} {\bibinfo  {journal} {J.
  Phys.: Condens. Matter}\ }\textbf {\bibinfo {volume} {7}},\ \bibinfo {pages}
  {759} (\bibinfo {year} {1995})}\BibitemShut {NoStop}%
\bibitem [{\citenamefont {Bocquet}\ and\ \citenamefont
  {Hill}(1995)}]{Bocquet_1995}%
  \BibitemOpen
  \bibfield  {author} {\bibinfo {author} {\bibfnamefont {S.}~\bibnamefont
  {Bocquet}}\ and\ \bibinfo {author} {\bibfnamefont {A.~J.}\ \bibnamefont
  {Hill}},\ }\href {\doibase 10.1007/BF00209379} {\bibfield  {journal}
  {\bibinfo  {journal} {Phys. Chem. Miner.}\ }\textbf {\bibinfo {volume}
  {22}},\ \bibinfo {pages} {524} (\bibinfo {year} {1995})}\BibitemShut
  {NoStop}%
\bibitem [{\citenamefont {Forsyth}\ \emph {et~al.}(1968)\citenamefont
  {Forsyth}, \citenamefont {Hedley},\ and\ \citenamefont
  {Johnson}}]{Forsyth_1968}%
  \BibitemOpen
  \bibfield  {author} {\bibinfo {author} {\bibfnamefont {J.~B.}\ \bibnamefont
  {Forsyth}}, \bibinfo {author} {\bibfnamefont {I.~G.}\ \bibnamefont {Hedley}},
  \ and\ \bibinfo {author} {\bibfnamefont {C.~E.}\ \bibnamefont {Johnson}},\
  }\href {\doibase 10.1088/0022-3719/1/1/321} {\bibfield  {journal} {\bibinfo
  {journal} {J. Phys. C: Solid State Phys.}\ }\textbf {\bibinfo {volume} {1}},\
  \bibinfo {pages} {179} (\bibinfo {year} {1968})}\BibitemShut {NoStop}%
\bibitem [{\citenamefont {Zepeda-Alarcon}\ \emph {et~al.}(2014)\citenamefont
  {Zepeda-Alarcon}, \citenamefont {Nakotte}, \citenamefont {Gualtieri},
  \citenamefont {King}, \citenamefont {Page}, \citenamefont {Vogel},
  \citenamefont {Wang},\ and\ \citenamefont {Wenk}}]{Zepeda_Alarcon_2014}%
  \BibitemOpen
  \bibfield  {author} {\bibinfo {author} {\bibfnamefont {E.}~\bibnamefont
  {Zepeda-Alarcon}}, \bibinfo {author} {\bibfnamefont {H.}~\bibnamefont
  {Nakotte}}, \bibinfo {author} {\bibfnamefont {A.~F.}\ \bibnamefont
  {Gualtieri}}, \bibinfo {author} {\bibfnamefont {G.}~\bibnamefont {King}},
  \bibinfo {author} {\bibfnamefont {K.}~\bibnamefont {Page}}, \bibinfo {author}
  {\bibfnamefont {S.~C.}\ \bibnamefont {Vogel}}, \bibinfo {author}
  {\bibfnamefont {H.-W.}\ \bibnamefont {Wang}}, \ and\ \bibinfo {author}
  {\bibfnamefont {H.-R.}\ \bibnamefont {Wenk}},\ }\href {\doibase
  10.1107/S1600576714022651} {\bibfield  {journal} {\bibinfo  {journal} {J.
  Appl. Cryst.}\ }\textbf {\bibinfo {volume} {47}},\ \bibinfo {pages} {1983}
  (\bibinfo {year} {2014})}\BibitemShut {NoStop}%
\bibitem [{\citenamefont {Kovalev}(1973)}]{Kovalev_1973}%
  \BibitemOpen
  \bibfield  {author} {\bibinfo {author} {\bibfnamefont {O.~V.}\ \bibnamefont
  {Kovalev}},\ }\href@noop {} {\bibfield  {journal} {\bibinfo  {journal} {Sov.
  Phys. Cryst.}\ }\textbf {\bibinfo {volume} {18}},\ \bibinfo {pages} {137}
  (\bibinfo {year} {1973})}\BibitemShut {NoStop}%
\bibitem [{\citenamefont {Sakhnenko}\ and\ \citenamefont
  {Ter-Oganessian}(2012)}]{Sakhnenko_2012}%
  \BibitemOpen
  \bibfield  {author} {\bibinfo {author} {\bibfnamefont {V.~P.}\ \bibnamefont
  {Sakhnenko}}\ and\ \bibinfo {author} {\bibfnamefont {N.~V.}\ \bibnamefont
  {Ter-Oganessian}},\ }\href {\doibase 10.1088/0953-8984/24/26/266002}
  {\bibfield  {journal} {\bibinfo  {journal} {J. Phys.: Condens. Matter}\
  }\textbf {\bibinfo {volume} {24}},\ \bibinfo {pages} {266002} (\bibinfo
  {year} {2012})}\BibitemShut {NoStop}%
\bibitem [{\citenamefont {Martin-Hernandez}\ and\ \citenamefont
  {Garc\'{i}a-Hern\'{a}ndez}(2010)}]{Martin_Hernandez_2010}%
  \BibitemOpen
  \bibfield  {author} {\bibinfo {author} {\bibfnamefont {F.}~\bibnamefont
  {Martin-Hernandez}}\ and\ \bibinfo {author} {\bibfnamefont {M.~M.}\
  \bibnamefont {Garc\'{i}a-Hern\'{a}ndez}},\ }\href {\doibase
  10.1111/j.1365-246X.2010.04566.x} {\bibfield  {journal} {\bibinfo  {journal}
  {Geophys. J. Int.}\ }\textbf {\bibinfo {volume} {181}},\ \bibinfo {pages}
  {756} (\bibinfo {year} {2010})}\BibitemShut {NoStop}%
\bibitem [{\citenamefont {Pankhurst}\ \emph {et~al.}(2012)\citenamefont
  {Pankhurst}, \citenamefont {Barqu\'{i}n}, \citenamefont {Lord}, \citenamefont
  {Amato},\ and\ \citenamefont {Zimmermann}}]{Pankhurst_2012}%
  \BibitemOpen
  \bibfield  {author} {\bibinfo {author} {\bibfnamefont {Q.~A.}\ \bibnamefont
  {Pankhurst}}, \bibinfo {author} {\bibfnamefont {L.~F.}\ \bibnamefont
  {Barqu\'{i}n}}, \bibinfo {author} {\bibfnamefont {J.~S.}\ \bibnamefont
  {Lord}}, \bibinfo {author} {\bibfnamefont {A.}~\bibnamefont {Amato}}, \ and\
  \bibinfo {author} {\bibfnamefont {U.}~\bibnamefont {Zimmermann}},\ }\href
  {\doibase 10.1103/PhysRevB.85.174437} {\bibfield  {journal} {\bibinfo
  {journal} {Phys. Rev. B}\ }\textbf {\bibinfo {volume} {85}},\ \bibinfo
  {pages} {174437} (\bibinfo {year} {2012})}\BibitemShut {NoStop}%
\bibitem [{\citenamefont {Kornev}\ \emph {et~al.}(2000)\citenamefont {Kornev},
  \citenamefont {Bichurin}, \citenamefont {Rivera}, \citenamefont {Gentil},
  \citenamefont {Schmid}, \citenamefont {Jansen},\ and\ \citenamefont
  {Wyder}}]{Kornev_2000}%
  \BibitemOpen
  \bibfield  {author} {\bibinfo {author} {\bibfnamefont {I.}~\bibnamefont
  {Kornev}}, \bibinfo {author} {\bibfnamefont {M.}~\bibnamefont {Bichurin}},
  \bibinfo {author} {\bibfnamefont {J.-P.}\ \bibnamefont {Rivera}}, \bibinfo
  {author} {\bibfnamefont {S.}~\bibnamefont {Gentil}}, \bibinfo {author}
  {\bibfnamefont {H.}~\bibnamefont {Schmid}}, \bibinfo {author} {\bibfnamefont
  {A.~G.~M.}\ \bibnamefont {Jansen}}, \ and\ \bibinfo {author} {\bibfnamefont
  {P.}~\bibnamefont {Wyder}},\ }\href {\doibase 10.1103/PhysRevB.62.12247}
  {\bibfield  {journal} {\bibinfo  {journal} {Phys. Rev. B}\ }\textbf {\bibinfo
  {volume} {62}},\ \bibinfo {pages} {12247} (\bibinfo {year}
  {2000})}\BibitemShut {NoStop}%
\bibitem [{\citenamefont {Yahia}\ \emph {et~al.}(2014)\citenamefont {Yahia},
  \citenamefont {Shikano}, \citenamefont {Tabuchi}, \citenamefont {Kobayashi},
  \citenamefont {Avdeev}, \citenamefont {Tan}, \citenamefont {Liu},\ and\
  \citenamefont {Ling}}]{Yahia_CoOHF_2014}%
  \BibitemOpen
  \bibfield  {author} {\bibinfo {author} {\bibfnamefont {H.~B.}\ \bibnamefont
  {Yahia}}, \bibinfo {author} {\bibfnamefont {M.}~\bibnamefont {Shikano}},
  \bibinfo {author} {\bibfnamefont {M.}~\bibnamefont {Tabuchi}}, \bibinfo
  {author} {\bibfnamefont {H.}~\bibnamefont {Kobayashi}}, \bibinfo {author}
  {\bibfnamefont {M.}~\bibnamefont {Avdeev}}, \bibinfo {author} {\bibfnamefont
  {T.~T.}\ \bibnamefont {Tan}}, \bibinfo {author} {\bibfnamefont
  {S.}~\bibnamefont {Liu}}, \ and\ \bibinfo {author} {\bibfnamefont {C.~D.}\
  \bibnamefont {Ling}},\ }\href {\doibase 10.1021/ic402294g} {\bibfield
  {journal} {\bibinfo  {journal} {Inorg. Chem.}\ }\textbf {\bibinfo {volume}
  {53}},\ \bibinfo {pages} {365} (\bibinfo {year} {2014})}\BibitemShut
  {NoStop}%
\bibitem [{\citenamefont {Kresse}\ and\ \citenamefont
  {Furthm\"{u}ller}(1996)}]{Kresse_1996}%
  \BibitemOpen
  \bibfield  {author} {\bibinfo {author} {\bibfnamefont {G.}~\bibnamefont
  {Kresse}}\ and\ \bibinfo {author} {\bibfnamefont {J.}~\bibnamefont
  {Furthm\"{u}ller}},\ }\href {\doibase 10.1103/PhysRevB.54.11169} {\bibfield
  {journal} {\bibinfo  {journal} {Phys. Rev. B}\ }\textbf {\bibinfo {volume}
  {54}},\ \bibinfo {pages} {11169} (\bibinfo {year} {1996})}\BibitemShut
  {NoStop}%
\bibitem [{\citenamefont {Bl\"{o}chl}(1994)}]{Bloechl_1994}%
  \BibitemOpen
  \bibfield  {author} {\bibinfo {author} {\bibfnamefont {P.~E.}\ \bibnamefont
  {Bl\"{o}chl}},\ }\href {\doibase 10.1103/PhysRevB.50.17953} {\bibfield
  {journal} {\bibinfo  {journal} {Phys. Rev. B}\ }\textbf {\bibinfo {volume}
  {50}},\ \bibinfo {pages} {17953} (\bibinfo {year} {1994})}\BibitemShut
  {NoStop}%
\bibitem [{\citenamefont {Dudarev}\ \emph {et~al.}(1998)\citenamefont
  {Dudarev}, \citenamefont {Botton}, \citenamefont {Savrasov}, \citenamefont
  {Humphreys},\ and\ \citenamefont {Sutton}}]{Dudarev_1998}%
  \BibitemOpen
  \bibfield  {author} {\bibinfo {author} {\bibfnamefont {S.~L.}\ \bibnamefont
  {Dudarev}}, \bibinfo {author} {\bibfnamefont {G.~A.}\ \bibnamefont {Botton}},
  \bibinfo {author} {\bibfnamefont {S.~Y.}\ \bibnamefont {Savrasov}}, \bibinfo
  {author} {\bibfnamefont {C.~J.}\ \bibnamefont {Humphreys}}, \ and\ \bibinfo
  {author} {\bibfnamefont {A.~P.}\ \bibnamefont {Sutton}},\ }\href {\doibase
  10.1103/PhysRevB.57.1505} {\bibfield  {journal} {\bibinfo  {journal} {Phys.
  Rev. B}\ }\textbf {\bibinfo {volume} {57}},\ \bibinfo {pages} {1505}
  (\bibinfo {year} {1998})}\BibitemShut {NoStop}%
\bibitem [{\citenamefont {Tunega}(2012)}]{Tunega_DFT_2012}%
  \BibitemOpen
  \bibfield  {author} {\bibinfo {author} {\bibfnamefont {D.}~\bibnamefont
  {Tunega}},\ }\href {\doibase 10.1021/jp2091297} {\bibfield  {journal}
  {\bibinfo  {journal} {J. Phys. Chem. C}\ }\textbf {\bibinfo {volume} {116}},\
  \bibinfo {pages} {6703} (\bibinfo {year} {2012})}\BibitemShut {NoStop}%
\bibitem [{\citenamefont {Meng}\ \emph {et~al.}(2016)\citenamefont {Meng},
  \citenamefont {Liu}, \citenamefont {Huo}, \citenamefont {Guo}, \citenamefont
  {Cao}, \citenamefont {Peng}, \citenamefont {Dearden}, \citenamefont {Gonze},
  \citenamefont {Yang}, \citenamefont {Wang}, \citenamefont {Jiao},
  \citenamefont {Li},\ and\ \citenamefont {Wen}}]{Meng_When_meets_oxides_2016}%
  \BibitemOpen
  \bibfield  {author} {\bibinfo {author} {\bibfnamefont {Y.}~\bibnamefont
  {Meng}}, \bibinfo {author} {\bibfnamefont {X.}~\bibnamefont {Liu}}, \bibinfo
  {author} {\bibfnamefont {C.}~\bibnamefont {Huo}}, \bibinfo {author}
  {\bibfnamefont {W.}~\bibnamefont {Guo}}, \bibinfo {author} {\bibfnamefont
  {D.}~\bibnamefont {Cao}}, \bibinfo {author} {\bibfnamefont {Q.}~\bibnamefont
  {Peng}}, \bibinfo {author} {\bibfnamefont {A.}~\bibnamefont {Dearden}},
  \bibinfo {author} {\bibfnamefont {X.}~\bibnamefont {Gonze}}, \bibinfo
  {author} {\bibfnamefont {Y.}~\bibnamefont {Yang}}, \bibinfo {author}
  {\bibfnamefont {J.}~\bibnamefont {Wang}}, \bibinfo {author} {\bibfnamefont
  {H.}~\bibnamefont {Jiao}}, \bibinfo {author} {\bibfnamefont {Y.}~\bibnamefont
  {Li}}, \ and\ \bibinfo {author} {\bibfnamefont {X.}~\bibnamefont {Wen}},\
  }\href {\doibase 10.1021/acs.jctc.6b00640} {\bibfield  {journal} {\bibinfo
  {journal} {J. Chem. Theory Comput.}\ }\textbf {\bibinfo {volume} {12}},\
  \bibinfo {pages} {5132} (\bibinfo {year} {2016})}\BibitemShut {NoStop}%
\bibitem [{\citenamefont {Bocquet}\ and\ \citenamefont
  {Kennedy}(1992)}]{Bocquet_1992}%
  \BibitemOpen
  \bibfield  {author} {\bibinfo {author} {\bibfnamefont {S.}~\bibnamefont
  {Bocquet}}\ and\ \bibinfo {author} {\bibfnamefont {S.~J.}\ \bibnamefont
  {Kennedy}},\ }\href {\doibase 10.1016/0304-8853(92)91758-L} {\bibfield
  {journal} {\bibinfo  {journal} {J. Magn. Magn. Mater.}\ }\textbf {\bibinfo
  {volume} {109}},\ \bibinfo {pages} {260} (\bibinfo {year}
  {1992})}\BibitemShut {NoStop}%
\bibitem [{\citenamefont {Leland}\ and\ \citenamefont
  {Bard}(1987)}]{Leland_1987}%
  \BibitemOpen
  \bibfield  {author} {\bibinfo {author} {\bibfnamefont {J.~K.}\ \bibnamefont
  {Leland}}\ and\ \bibinfo {author} {\bibfnamefont {A.~J.}\ \bibnamefont
  {Bard}},\ }\href {\doibase 10.1021/j100303a039} {\bibfield  {journal}
  {\bibinfo  {journal} {J. Phys. Chem.}\ }\textbf {\bibinfo {volume} {91}},\
  \bibinfo {pages} {5076} (\bibinfo {year} {1987})}\BibitemShut {NoStop}%
\bibitem [{\citenamefont {Sherman}(2005)}]{Sherman_2005}%
  \BibitemOpen
  \bibfield  {author} {\bibinfo {author} {\bibfnamefont {D.~M.}\ \bibnamefont
  {Sherman}},\ }\href {\doibase 10.1016/j.gca.2005.01.023} {\bibfield
  {journal} {\bibinfo  {journal} {Geochim. Cosmochim. Acta}\ }\textbf {\bibinfo
  {volume} {69}},\ \bibinfo {pages} {3249} (\bibinfo {year}
  {2005})}\BibitemShut {NoStop}%
\end{thebibliography}


%

\end{document}